\documentclass[12pt]{iopart}
\usepackage{enumerate}
\usepackage[section]{placeins}
\expandafter\let\csname equation*\endcsname\relax
\expandafter\let\csname endequation*\endcsname\relax

\usepackage{amsmath,amsfonts,amssymb}
\usepackage[english]{babel}

\let\p\partial

\let\ds\displaystyle

\newcommand{\A}[0]{{\cal A}}

\newcounter{lemma}
\newcommand{\lemma}[1]{\vskip 1mm{\bf Lemma \arabic{lemma}. \stepcounter{lemma}}{\it #1}\vskip 1mm}

\newcounter{theorem}
\newcommand{\theorem}[1]{\vskip 1mm{\bf Theorem \arabic{theorem}. \stepcounter{theorem}}{\it #1}\vskip 1mm}

\newcounter{definition}

\begin{document}

\title{Darboux integrability of determinant and equations for principal minors} %

\author{D K Demskoi${}^1$, D T  Tran${}^2$}
\address{
${}^1$ School of Computing and Mathematics,\\ Charles Sturt University, NSW 2678, Australia \\[2mm]
${}^2$ Department of Mathematics and Statistics,\\ La Trobe University, Victoria 3086, Australia
}

\begin{abstract}
We consider equations that represent a constancy condition for a 2D Wronskian, mixed Wronskian-Casoratian and 2D Casoratian. 
These determinantal equations are shown to have the number of independent integrals equal to their order - this implies Darboux integrability.
On the other hand, the recurrent formulas for the leading principal minors 
are equivalent to the 2D Toda equation and its semi-discrete and lattice analogues with particular boundary conditions (cut-off constraints).
This connection is used to obtain recurrent formulas and closed-form expressions for integrals of the Toda-type equations
from the integrals of the determinantal equations. 
General solutions of the equations corresponding to vanishing determinants are given explicitly 
while in the non-vanishing case they are given in terms of solutions of ordinary linear equations.
\end{abstract}

\section{Introduction}
Consider the scalar equation
\begin{equation}
	\det \A=\beta,	
	\label{det1}
\end{equation}
where $\beta=\mbox{const}$ and the entries $a_{ij}$ of matrix $\A$ have one of the  forms
$$
a_{i+1,j+1}=\frac{\p^{i+j}}{\p x^i \p t^j}u(t,x), \ \ a_{i+1,j+1}=\frac{d^j}{d t^j}u(t,k+i),\ \ a_{i+1,j+1}=u(k+i,l+j)
$$
with $i,j=0,\dots,N-1$. The independent variables $t$ and $x$ can be either real or complex, while $k$ and $l$ are integers. 
These three forms give rise to correspondingly the partial differential, semi-discrete (differential-difference) and lattice equations.
The left-hand side of (\ref{det1}) in the continuous case is sometimes referred to as a 2D Wronskian, since it can be interpreted as either
of the Wronskians: $W(u,u_t,\dots,u_{t\dots t})(x)$ or $W(u,u_x,\dots,u_{x\dots x})(t)$. Similarly, in the semi-discrete case it
can be regarded as either the Wronskian $W(u_k,\dots, u_{k+N-1})(t)$ or the Casoratian 
$C(u,  u^{(1)},\dots, u^{(N-1)})(k)$. The lattice case gives rise to an object we will refer to as a 2D Casoratian.
These objects naturally arise in construction of solutions of Toda-type equations, both continuous and discrete (see e.g. \cite{hir0}-\cite{hir2}).

Apart from scalar equation (\ref{det1}), we are concerned with systems of equations that can be derived from (\ref{det1}).
We are particularly interested in the system obtained by denoting $w_n$ as the leading principal minors of matrix $\A$.
The first publications on continuous case date back to Darboux \cite{darb}, who found that $w_n$ satisfies the chain of equations
\begin{equation}
	\ds \p^2_{tx} \ln w_n=\frac{w_{n+1}w_{n-1}}{w_n^2},\ \ n=1\dots N-1
	\label{todaln}
	\end{equation}
with the boundary conditions
\begin{eqnarray}
	w_0=1,\label{ws0} \\ 
	w_N= \beta,\label{ws}
\end{eqnarray}
where (\ref{ws}) is simply another representation of equation (\ref{det1}).
System (\ref{todaln}) is equivalent to a two-dimensional (2D) Toda equation 
which arguably received more attention than equation (\ref{det1}) itself. 
Nevertheless equation (\ref{det1}) is, due to its transparent algebraic
structure, very convenient for studying certain properties of system (\ref{todaln}). 
For instance, the {\it local} integrals of (\ref{det1}) have simple expression in terms of minors of matrix $\A$.
The integrals of system (\ref{todaln}) and its reductions can then be derived from the integrals of the scalar equation
as demonstrated in \cite{demint} for the continuous case. In this paper we elaborate on this result deriving integrals in a similar fashion
for the semi-discrete and lattice cases. For completeness of exposition we also give proofs in the continuous case. 

Note that presence of integrals entirely depends on boundary conditions (\ref{ws0}) and (\ref{ws}).
Imposing different boundary conditions, for instance the periodic condition $w_{n+i}=w_n$, drastically changes the properties
of (\ref{todaln}). It is well-known \cite{fordy} that 2D Toda equations with 
the periodic condition (for $i>1$) yield solitonic equations. 
On the other hand, non-linear equations that possess local integrals form a subclass of linearisable equations. 
The famous example of such an equation is the Liouville equation
\begin{equation}
	v_{tx}=-\beta \exp(2 v).
	\label{liouville}
\end{equation}
Indeed, one can easily verify that if $z$ solves the d'Alembert equation $z_{tx}=0$, then
$$
v=\frac{1}{2}\ln\frac{z_xz_t}{-\beta z^2}
$$
is the solution of the Liouville equation.
Note that (\ref{liouville}) can be written in form (\ref{det1}) if we substitute $v=-\ln u.$
Sometimes the whole class of equations admitting local integrals is referred to as the Liouville-type equations. 
The other commonly used name is the {\it Darboux integrable} equations. 
There is a vast body of knowledge on these equations, including classification results (see e.g. \cite{gours,vessiot,zhsok,anders} and references therein). 
Some progress has been also achieved in the study of systems of equations admitting integrals, especially the exponential systems
associated with simple Lie algebras \cite{shabyam,shablez,leznov,zhgur,nie}.

The study of discrete analogues of the Liouville equation 
started relatively recently \cite{adst}. The main feature of such equations is the
presence of local integrals just as it is in the continuous case. It has been shown that these equations also have terminating sequences of the Laplace invariants \cite{adst}
and finite-dimensional characteristic algebras \cite{habib0}. 
Discrete analogues of the 2D Toda equation corresponding to various Cartan matrices were investigated in \cite{kh1,habib1}.
A general construction of integrals for semi-discrete and lattice Toda systems of type $A$ and $C$ was proposed in \cite{smirnov}.
A different approach based on the consistency properties of a particular case of scalar equation (\ref{det1}), namely the lattice equation 
with $\beta=0$, was introduced in \cite{mokhov}. The consistency approach (see e.g. \cite{nijh1,bob1,bob2}) was also applied to study equations associated with
hyper-determinants \cite{tsar1}.


In this paper equation (\ref{det1}) is studied from the view-point of Darboux integrability. Our main objective
is construction of integrals and general solutions of this equation. The result is then transferred to the case of the semi-discrete and lattice
analogues of the Toda equation with boundary conditions (\ref{ws0}) and (\ref{ws}). 

The rest of the paper is organized as follows.
Firstly, we consider in detail derivation of integrals for the Liouville equation and its differential-difference
analogue. The derivation of integrals in the lattice case is similar to the semi-discrete one. The example of the Liouville equation 
also illustrates our approach to integrals of equation (\ref{det1}) which is developed in the following sections. 
The concluding section contains the outline of a procedure for construction of general solutions.

\section{Integrals of the Liouville equation}\label{Liouv}

The Liouville equation has a remarkable property that it
possesses local functions of solutions and their derivatives, let us denote them as $I(v,v_t,\dots)$ and $J(v,v_x,\dots)$, 
that depend only on one variable along the characteristics:
\begin{equation}
	\p_x I=0,\ \ \p_t J=0.
	\label{intzero}
\end{equation}
These functions are called the $x$- and $t$-integrals. Note that $I$ and $J$ do not depend on $x$- and $t$-derivatives 
correspondingly. This is a rather general property (see e.g. \cite{zhsok}) of hyperbolic equations admitting integrals.
The simplest integrals of the Liouville equation have the form
\begin{equation}
	I=v_{tt}-v_t^2, \ \ J=v_{xx}-v_x^2.
	\label{liouvint}
\end{equation}
All other integrals can be shown to be functions of $I$ and $J$ and their derivatives.
The  fulfillment of (\ref{intzero}) is verified by differentiating the integrals 
and eliminating the mixed derivatives by means of (\ref{liouville}). 
In relation to integrals we are also interested in a particular representation of their derivatives
\begin{equation}
	\p_x I=\Lambda_I F,\ \ \p_t J=\Lambda_J F
	\label{chargen}
\end{equation}
which is reminiscent of characteristic form of conservation laws. Here $\Lambda_I$ and $\Lambda_J$
are some differential operators, which we call {\it integrating factors}, and $F$ represents the equation in question.
For example, the derivatives 
of (\ref{liouvint}) can be written as
\begin{equation}
	\p_x I=(\p_t-2 v_t)(v_{tx}+\beta e^{2v}),\ \ \p_t J=(\p_x-2 v_x)(v_{tx}+\beta e^{2v}).
	\label{liouvchar}
\end{equation}
For lattice equations the analog of (\ref{chargen}) is the representation
\begin{equation}
	S_k(I)-I=\Lambda_I F,\ \ S_l(J)-J=\Lambda_J F,
	\label{chargend}
\end{equation}
where $S_k$ and $S_l$ are the shift operators with respect to the subscripted variables. In the semi-discrete
case we have two types of integrals hence the mixture of (\ref{chargen}) and (\ref{chargend}):
\begin{equation}
	S_k(I)-I=\Lambda_I F,\ \ \p_t J=\Lambda_J F.
		\label{chargensd}
\end{equation}
Our interest in representation (\ref{chargen}) 
is motivated by the fact that it contains important information about the structure of co-symmetries and higher symmetries
of the equation. 
For example, for the Liouville equation it can be shown that the formal adjoint operators
$$
\Lambda_I^{T}=-\p_t-2 v_t,\ \ \Lambda_J^{T}=-\p_x-2 v_x
$$
map correspondingly functions of integrals $I$ and $J$ to symmetries of (\ref{liouville}). 
For example, the simplest higher symmetry $v_\tau=\Lambda_J^{T} J$ is the mKdV equation in the potential form
$$
v_\tau=-v_{xxx}+2 v_x^3.
$$
A detailed exposition on construction of symmetries from integrals of hyperbolic equations is discussed in \cite{demstar}.
The representation of the form (\ref{chargend}) was used to introduce the notion of {\it dual equation} in \cite{robquisp}.
In this paper we restrict ourselves to finding representations (\ref{chargen}), (\ref{chargensd}) and (\ref{chargend}). The problem
of construction of higher symmetries from the integrals will be considered elsewhere.

Now we describe how the integrals of the Liouville equation and its discrete counterparts can be obtained from determinantal form (\ref{det1}).
As it has been pointed out already, the Liouville equation is equivalent to (\ref{det1}) for $N=2$:
\begin{equation}
\begin{vmatrix}
u & u_t \\
u_x & u_{tx}\\
\end{vmatrix}=\beta.
\label{w3}	
\end{equation}
Differentiating (\ref{w3}) with respect to $x$ we obtain 
\begin{equation}
\begin{vmatrix}
u & u_{t} \\
u_{xx} & u_{xxt} \\
\end{vmatrix}=0.
\end{equation}
Hence the rows of the determinant are proportional:
$$
(u_{xx} , u_{xxt} )=J(x)(u , u_t)
$$
and the coefficient of proportionality 
\begin{equation}
	J=\frac{u_{xx}}{u}
	\label{liouvint0}
\end{equation}
is a $t-$integral of equation (\ref{w3}). Representation $(\ref{chargen})_2$ is then easily found
$$
\p_t J=\frac{1}{u^2}\p_x\left(\begin{vmatrix}
u & u_t \\
u_x & u_{tx}\\
\end{vmatrix}-\beta\right).
$$
Integral (\ref{liouvint0}) takes form $(\ref{liouvint})_2$ on substituting $u=\exp(-v)$.
The $t-$integral is obtained by differentiating (\ref{w3}) with respect to $t$ and finding
the coefficient of proportionality of columns. Obviously, it is the same as (\ref{liouvint0}) with $x$-
and $t$-derivatives interchanged.

\subsubsection*{Semi-discrete Liouville equation}

We use the determinant form of the Liouville equation to introduce its differential-difference and lattice analogues 
by simply replacing the derivatives with shifts. Let us consider the differential-difference equation
\begin{equation}
\begin{vmatrix}
u_k & \dot{u}_k \\[1mm]
u_{k+1} & \dot{u}_{k+1}\\
\end{vmatrix}=\beta,
\label{w2dd}	
\end{equation}
where the ``dot'' indicates the $t$-derivative.
Naming (\ref{w2dd}) as a differential-difference Liouville equation is justified because it has (\ref{w3}) as its continuum limit.
Indeed, we obtain (\ref{w3}) on setting $x=k\varepsilon,\, u_k(t)=u(t,x)/\sqrt{\varepsilon}$ in (\ref{w2dd}) and passing to the limit 
$\varepsilon\to 0$. Other differential-difference and lattice avatars of the Liouville equation can be found in \cite{adst,rebel,winter}.

On differentiating (\ref{w2dd}) we obtain
\begin{equation}
\begin{vmatrix}
u_k & \ddot{u}_k \\[1mm]
u_{k+1} & \ddot{u}_{k+1}\\
\end{vmatrix}=0
\label{w2ddd}	
\end{equation}
which implies the relation 
$$(\ddot{u}_k,\ddot{u}_{k+1})=I(t)(u_k,u_{k+1}).$$ 
The coefficient of proportionality
\begin{equation}
	I=\frac{\ddot{u}_k}{u_k}
	\label{liouvintsd}
\end{equation}
is the $k$-integral of (\ref{w2dd}). Differencing (\ref{liouvintsd}) we find the representation
$$
\Delta_k I=\frac{1}{u_k u_{k+1}}\frac{d}{dt}\bigg(\begin{vmatrix}
u_k & \dot{u}_k \\[1mm]
u_{k+1} & \dot{u}_{k+1}\\
\end{vmatrix}-\beta\bigg),
$$
where $$\Delta_k=S_k-1.$$

In order to find the $t$-integral of (\ref{w2dd}) we difference the equation to obtain
$$
\begin{vmatrix}
u_{k+1} & \dot{u}_{k+1} \\[1mm]
u_{k+2} & \dot{u}_{k+2}\\
\end{vmatrix}-\begin{vmatrix}
u_k & \dot{u}_k \\[1mm]
u_{k+1} & \dot{u}_{k+1}\\
\end{vmatrix}=0
$$
which can be written as the single determinant
\begin{equation}
\begin{vmatrix}
u_{k+1} & \dot{u}_{k+1} \\[1mm]
u_{k+2}+u_k & \dot{u}_{k+2}+\dot{u}_k\\
\end{vmatrix}=0.	
\end{equation}
The latter implies 
$$
(u_{k+2}+u_k , \dot{u}_{k+2}+\dot{u}_k)=J(k)(u_{k+1} , \dot{u}_{k+1})
$$ 
and the following form of the $t-$integral:
$$
J=\frac{u_{k+2}+u_k}{u_{k+1}}.
$$
One can verify that
$$
\dot{J}=\frac{1}{u_{k+1}^2}(S_k-1)\bigg(\begin{vmatrix}
u_k & \dot{u}_k \\[1mm]
u_{k+1} & \dot{u}_{k+1}\\
\end{vmatrix}-\beta\bigg)
$$
which shows that $\dot{J}=0$ on solutions of (\ref{w2dd}).

The above illustrates that in the differential-difference case we deal with the two different types of integrals.
Due to symmetries $t \leftrightarrow x$ and $k \leftrightarrow l$ in  
continuous and lattice cases, it is sufficient to indicate integrals with
respect to one variable only.
\section{Integrals for vanishing determinant}
\label{vanint}
In what follows Sylvester's determinant identity \cite{sylvester} plays an important role. 
Let $\A_n$ be the matrix obtained from $\A$ by retaining its first $n$ rows and columns and deleting the rest. 
According to the notation introduced earlier, $\det \A_n=w_n$.
Further, we denote the minor of the entry in the $p$-th row and $q$-th column of matrix $\A_n$ as $m_{pq}$ 
and the minor obtained from $\A_n$ by removing its $p$- and $q$-th rows as well as $r$- and $s$-th columns as $m_{pqrs}$. 
A particular case of Sylvester's identity (also known as Jacobi identity \cite{hirot}), that we require, reads
\begin{equation}
	m_{pqrs} w_n=m_{pr}m_{qs}-m_{ps}m_{qr}. 
	\label{sylvester}
\end{equation}
To indicate the order of minors $m_{pq}$ and $m_{pqrs}$ explicitly, we write $m_{n;pq}$ and $m_{n;pqrs}$ respectively.
This notation means that the minors are obtained by deleting rows and columns from matrix $\A_n$.
We will refer to particular instances of Sylvester's identity by indicating the list of indices $(p,q,r,s)$.
%

First we develop a procedure for construction of integrals of a vanishing determinant and then show (Section \ref{nonvan}) that 
it can be easily extended to the non-vanishing case similarly to how it is done for the Liouville equation in the previous section.

Vanishing of determinant $w_{N}$ implies a linear dependence of its columns or rows. Hence 
the last column and row can be decomposed correspondingly as
\begin{equation}
	a_{iN}=\sum_{r=1}^{N-1} a_{ir} I_{{N};r}, \ \ a_{{N}i}=\sum_{p=1}^{N-1} a_{pi} J_{{N};p}.
	\label{decomps}
\end{equation}
Clearly the coefficients $I_{N;r}$ do not depend on the variable that labels
different rows of the determinant. Functions $I_{N;r}$ are therefore the $x$-integrals in the continuous case
and $k$-integrals in the semi-discrete and lattice cases. Likewise the coefficients $J_{N;p}$ 
are the $t$-integrals in the continuous and semi-discrete cases and $l$-integrals in the lattice
case. 
Setting $i=1,\dots,N-1$ in (\ref{decomps}) and solving for $I_{N;r}$ and $J_{N;p}$ using Cramer's rule
we get explicit expressions for the integrals of (\ref{det1}):
\begin{equation}
	I_{N;r}=\frac{m_{N r}}{w_{N-1}},\ \ J_{N;p}=\frac{m_{pN}}{w_{N-1}},\ \ p,r=1,\dots,N-1.
	\label{intskt}
\end{equation}
For consistency of formulas that follow we also define
\begin{equation}
	I_{N;N}=J_{N;N}=1, \ \ I_{N;0}=J_{N;0}=0.
	\label{sl0}
\end{equation}

{\bf Remark.} It is easy to see that homogeneous polynomials $m_{N r}$ are linearly independent and of the same order $2(N-1)$. 
On the other hand, each of these polynomials depends on a unique set of variables.
This implies functional independence of integrals $I_{N;r}$. Apparently the same argument applies to integrals
$J_{N;p}$.

Below we show that functions $I_{n;r}$ and $J_{n;p}$ satisfy some remarkable identities which imply
representations of form (\ref{chargen}), (\ref{chargend})  and (\ref{chargensd}).
%
\stepcounter{lemma}
\lemma{ The functions $I_{n;r}$ and $J_{n;p}$ satisfy the following identities.
\begin{enumerate}[]
\item
Continuous case:
\begin{equation}
	\p_x I_{n;r}=\frac{w_{n-2}w_n}{w_{n-1}^2}I_{n-1;r},\ \ \p_t J_{n;p}=\frac{w_{n-2}w_n}{w_{n-1}^2}J_{n-1;p}.
	\label{jcont}
\end{equation}
\item
Semi-discrete case:
\begin{equation}
	\Delta_k I_{n;r}=\frac{w_n}{w_{n-1}}S_k\Big(\frac{w_{n-2}}{w_{n-1}}I_{n-1;r}\Big), \ \ \frac{d J_{n;p}}{dt}=\frac{w_{n-2}w_n}{w_{n-1}^2}J_{n-1;p}.
\label{IJcontd}
\end{equation}
\item
Lattice case:
\begin{equation}
\Delta_k I_{n;r}=\frac{w_n}{w_{n-1}}S_k\Big(\frac{w_{n-2}}{w_{n-1}}I_{n-1;r}\Big) ,\ \ 	\Delta_l J_{n;p}=\frac{w_n}{w_{n-1}}S_l\Big(\frac{w_{n-2}}{w_{n-1}}J_{n-1;p}\Big).
	\label{ddints}
\end{equation}
\end{enumerate}
}
{\bf Proof.} 
The idea of the proof is to match the derivative (difference) of $I_{n;r}$ and $J_{n;p}$
with the right-hand-side of Sylvester's identity for some
$(p,q,r,s)$.
We consider the semi-discrete case in detail as it is illustrative of both the continuous and lattice cases. 

Note that the difference 
$$
\Delta_k I_{n;r}=S_k\Big(\frac{m_{nr}}{w_{n-1}}\Big)-\frac{m_{nr}}{w_{n-1}}$$ 
can be written as
\begin{equation}
	\Delta_k I_{n;r}=\frac{1}{w_{n-1}S_k(w_{n-1})}\big(S_k(m_{nr})w_{n-1}-S_k(w_{n-1})m_{nr}\big).
	\label{proofintsd}
\end{equation}
Taking into account the relations 
$$ 
S_k(m_{nr})=m_{1r},\ \ w_{n-1}=m_{nn},\ \ S_k(w_{n-1})=m_{1n},\ \  m_{n;1nrn}=m_{n-1;1r},
$$
and using Sylvester's identity with $(1,n,r,n)$ we re-write the expression in brackets as
$$
\begin{array}{ll}
S_k(m_{nr})w_{n-1}-S_k(w_{n-1})m_{nr}&= m_{1r}m_{nn}-m_{1n}m_{nr}\\[1mm]
& = m_{n-1;1r}w_n.
\end{array}
$$
After multiplying and dividing the right hand side of (\ref{proofintsd}) by $S_k(w_{n-2})$
and taking into account that $m_{n-1;1r}=S_k(m_{n-1;n-1,r})$ we obtain formula $(\ref{IJcontd})_1$.

Further, in the expression for the derivative
$$
\frac{dJ_{n;p}}{dt}=\frac{1}{w_{n-1}^2}\big(\dot{m}_{pn}w_{n-1}-m_{pn}\dot{w}_{n-1}\big)
$$
one can easily recognise that
$$
\dot{m}_{pn}=m_{p,n-1},\ \ w_{n-1}=m_{nn},\ \ \dot{w}_{n-1}=m_{n,n-1},\ \ m_{n;pn,n-1,n}=m_{n-1;p,n-1}.
$$
Substituting the latter in the former, and taking into account Sylvester's identity with $(p,n,n-1,n)$, we obtain
$$
\frac{dJ_{n;p}}{dt} =\frac{w_n m_{n-1;p,n-1}}{w_{n-1}^2}
$$
and hence $(\ref{IJcontd})_2$.

The pairs of formulas $(\ref{jcont})_2$, $(\ref{IJcontd})_2$ and $(\ref{IJcontd})_1$, $(\ref{ddints})_1$
are identical, so are their proofs.


\section{Equations for leading principal minors}
\subsection{Derivation of equations}
In this section we derive the equations, or more correctly, the chains of equations
satisfied by the leading principal minors of matrix $\A_n$. In the continuous case we get
a version of the 2D Toda equation given by (\ref{todaln}). It is natural to expect that
in semi-discrete and lattice cases we get some analogs of the 2D Toda equation. 
The derivation relies on identities (\ref{jcont})-(\ref{ddints}) for functions $I$ and $J$.
\stepcounter{theorem}
\theorem{Quantities $w_n$ satisfy the following relations.
\begin{enumerate}[]
\item
Continuous case:
\begin{equation}
		\p^2_{tx} \ln w_n=\frac{w_{n-1}w_{n+1}}{w_n^2}. \label{2dtodac}
\end{equation}
\item
Semi-discrete case:
\begin{equation}
		\ds\Delta_k\frac{d}{dt} \ln w_{n}=\frac{w_{n+1}}{w_n}S_k\Big(\frac{w_{n-1}}{w_n}\Big). \label{2dtodacd} 
\end{equation}
\item
Lattice case:
\begin{equation}
		\ds\Delta_l \frac{w_{n}}{S_k(w_{n})}=-\frac{w_{n+1}}{S_k(w_{n})}S^2_{kl}\Big(\frac{w_{n-1}}{w_n}\Big).\label{2dtodacdd}
\end{equation}
\end{enumerate}
	
}
{\bf Proof.} 
%
The formula in the continuous case can be derived from $(\ref{jcont})_2$ if we substitute $p=n-1$
and notice that
$$
J_{n;n-1}=\p_x \ln w_{n-1},\ \ J_{n-1,n-1}=1.
$$
This results in down-shifted formula (\ref{2dtodac}).

Similarly, in the semi-discrete case we set $r=n-1$ in $(\ref{IJcontd})_1$ to obtain
$$
\Delta_k I_{n;n-1}=\frac{w_n }{w_{n-1}}S_k\Big(\frac{w_{n-2}}{w_{n-1}}\Big).
$$
This relation combined with
$$
I_{n;n-1}=\frac{d}{dt} \ln w_{n-1}
$$ 
gives formula (\ref{2dtodacd}) down-shifted with respect to $n$.

Equation (\ref{2dtodacdd}) is in fact equivalent to Sylvester's identity 
for the corner entries of matrix $\A_n$. Indeed, setting $(p,q,r,s)=(1,n,1,n)$
in (\ref{sylvester}) we obtain
$$
w_n m_{1n1n}=m_{11}m_{nn}-m_{1n}m_{n1}
$$
which can be re-written as
\begin{equation}
	w_n S^2_{kl}(w_{n-2})=S^2_{kl}(w_{n-1})w_{n-1}-S_k(w_{n-1}) S_l(w_{n-1}).
	\label{lastlattice}
\end{equation}
Equation (\ref{lastlattice}) is a different way to write (\ref{2dtodacdd}). Indeed, if we divide
(\ref{lastlattice}) by $S_{kl}^2(w_{n-1})S_k(w_{n-1})$ and recognise that the right-hand side of the 
resulting expression is the difference 
$$
(1-S_l)\frac{w_{n-1}}{S_k(w_{n-1})}=-\Delta_l\frac{w_{n-1}}{S_k(w_{n-1})}
$$
we obtain (\ref{2dtodacdd}) down-shifted with respect to $n$. This concludes the proof.

{\bf Remark.} Equation (\ref{2dtodacd}) with boundary conditions (\ref{ws0}) and (\ref{ws}) is equivalent (up to a simple transformation) to the  system 
introduced in \cite{kh1} as a discretisation of the classical 2D Toda system of type $A_N$
while equation (\ref{lastlattice}) is related to the Hirota equation \cite{hirot2}. 
%

\stepcounter{theorem}
\theorem{The quantities $I_{N;r},\, J_{N;p}$ 
are the integrals of equation (\ref{det1}) or equivalently of (\ref{2dtodac})-(\ref{2dtodacdd}) with boundary conditions (\ref{ws0}),(\ref{ws}), assuming $\beta=0$.
}
On substituting $n=N$ and $w_N=0$ to (\ref{jcont})-(\ref{ddints}) we note that the right-hand sides of these relations vanish which concludes the proof.


\subsection{Recurrent formulas for integrals}
We have shown that the formulas for integrals obtained in Section \ref{vanint} are suitable not only for scalar equation 
(\ref{det1}) but also for recurrences (\ref{2dtodac})-(\ref{2dtodacdd}) with the boundary conditions
(\ref{ws0}), (\ref{ws}). However, the integrals contain mixed derivatives (shifts) that can be eliminated by means of the
recurrences themselves. This would give a more natural expression for the integrals as they are usually given
in terms of derivatives (shifts) with respect to one variable only. Therefore our objective in this section
is to express the integrals $I$ and $J$ in terms of derivatives (shifts) of leading principal 
minors $w_n$. 

\theorem{Quantities $I_{n;r}$ and $J_{n;p}$ satisfy the following recurrent formulas.
\begin{enumerate}[]
\item Continuous case:
\begin{equation}
	J_{n;p}=J_{n-1;p-1}-\p_x J_{n-1;p}+J_{n-1;p}\p_x \ln\frac{w_{n-1}}{w_{n-2}}.
	\label{recurrent}
\end{equation}
The recurrent formula for $x-$integrals $I_{n;r}$ is identical to (\ref{recurrent}) with $x$ replaced by $t$.
\item Semi-discrete case:
\begin{equation}
		I_{n;r}=I_{n-1;r-1}-\dot{I}_{n-1,r}+I_{n-1;r}\frac{d}{dt}\ln \frac{w_{n-1}}{w_{n-2}},
		\label{recursd1}	
\end{equation}
\begin{equation}
		J_{n;p}=S_k(J_{n-1;p-1})+\frac{w_{n-2}}{w_{n-1}}S_k\Big(\frac{w_{n-1}}{w_{n-2}}\Big)J_{n-1;p}.
				\label{recusdt}
\end{equation}
\item Lattice case:
\begin{equation}
J_{n;p}=S_k(J_{n-1;p-1})+\frac{w_{n-2}}{w_{n-1}}S_k\Big(\frac{w_{n-1}}{w_{n-2}}\Big)J_{n-1;p}.
\label{laticerec}
\end{equation}	
The recurrent formula for $k-$integrals $I_{n;r}$ is identical to (\ref{laticerec}) with $k$ replaced by $l$.
\end{enumerate}
The initial and boundary conditions for these formulae are given by (cf. (\ref{sl0}))
\begin{equation}
		J_{1;0}=0,\ \ J_{1;1}=1, \ \ J_{n,0}=0, \ \ J_{n,n}=1.
		\label{iboundary}
\end{equation}
}
\subsubsection*{Proof. Continuous case.} 
We start with a relation that follows directly from the rule of differentiation of Wronskians:
\begin{equation}
m_{n-1;p-1,n-1}=\p_x m_{n-1;p,n-1}-m_{p,n-1,n-1,n}.
\label{initeq}
\end{equation}
Its last term can be re-written by means of Silvester's identity the following way
$$
\begin{array}{ll}
w_n m_{p,n-1,n-1,n}&=m_{p,n-1}m_{n-1,n}-m_{pn}m_{n-1,n-1} \\[2mm]
 &=\p_t m_{pn}\p_x w_{n-1}-m_{pn}\p_{tx}^2 w_{n-1}.
\end{array}
$$
Using (\ref{2dtodac}) and $(\ref{jcont})_2$ written correspondingly in the form 
$$
\p_{tx}^2 w_{n-1}=\frac{\p_t w_{n-1}\p_x w_{n-1}}{w_{n-1}}+\frac{w_{n-2} w_{n}}{w_{n-1}}, \ \ \p_t m_{pn}=\frac{m_{pn}\p_t w_{n-1}}{w_{n-1}}+\frac{m_{n-1;p,n-1}w_n}{w_{n-1}}
$$
we obtain
\begin{equation}
	m_{p,n-1,n-1,n}=\frac{\p_x w_{n-1}m_{n-1;p,n-1}}{w_{n-1}}-\frac{w_{n-2}m_{pn}}{w_{n-1}}.
\end{equation}
Substituting this formula in (\ref{initeq}) and dividing by $w_{n-2}$, we obtain the formula
\begin{equation}
	\frac{m_{pn}}{w_{n-1}}=\frac{m_{n-1;p,n-1}\p_x w_{n-1}}{w_{n-2}w_{n-1}}-\frac{\p_x m_{n-1;p,n-1}}{w_{n-2}}+\frac{m_{n-1;p-1,n-1}}{w_{n-2}}
	\label{prereq}
\end{equation}
equivalent to (\ref{recurrent}).

\subsubsection*{Semi-discrete case.}
Here we aim to obtain recurrent formulas that yield derivative-free $t$-integrals and shift-free $k$-integrals.
The constructions for $k$- and $t$-integrals appear to be slightly different, so we consider both in detail. 

{\it Formula for $k$-integrals.}
We start with the relation 
$$
m_{n-1;n-1,r}=\dot{m}_{n-1;n-1,r+1}-m_{n-1,n,r+1,n-1}
$$
which is identical to (\ref{initeq}). Its up-shifted version is given by 
\begin{equation}
	S_k(m_{n-1;n-1,r})=S_k(\dot{m}_{n-1;n-1,r+1})-m_{1,n,r+1,n-1}.
	\label{smnr}
\end{equation}
Consider Sylvester's identity with $(1,n,r+1,n-1)$:
\begin{equation}
	w_{n}m_{1,n,r+1,n-1}=m_{1,r+1}m_{n,n-1}-m_{1,n-1}m_{n,r+1}.
	\label{sylv1}
\end{equation}
The terms in the right-hand side of (\ref{sylv1}) can be interpreted as
$$
m_{1,r+1}=S_k(m_{n,r+1}), \ \ m_{n,n-1}=\dot{w}_{n-1},\ \ m_{1,n-1}=S_k(\dot{w}_{n-1} ).
$$
Substituting these expressions in (\ref{sylv1}) and also multiplying and dividing its right-hand side by $w_{n-1} S_k(w_{n-1})$, we obtain
\begin{equation}
	w_{n}m_{1,n,r+1,n-1}=w_{n-1}S_k(w_{n-1})\Big(S_k(I_{n;r+1})\frac{d}{dt}\ln w_{n-1}-I_{n;r+1}S_k\big(\frac{d}{dt}\ln w_{n-1}\big)\Big).
	\label{intermexp}
\end{equation}
Formula (\ref{intermexp}) can be simplified by means of the relations
$$
\begin{array}{l}
\ds I_{n;r+1}=S_k(I_{n;r+1})-\frac{w_{n}}{w_{n-1}}S_k\Big(\frac{w_{n-2}}{w_{n-1}}I_{n-1;r+1}\Big), \\[3mm]
\ds \frac{d}{dt}\ln w_{n-1}=S_k\Big(\frac{d}{dt}\ln w_{n-1}\Big)-\frac{w_{n}}{w_{n-1}}S_k\Big(\frac{w_{n-2}}{w_{n-1}}\Big)
\end{array}
$$
which follow from $(\ref{IJcontd})_1$ and $(\ref{2dtodacd})_1$ correspondingly. The formula then becomes
\begin{equation}
	m_{1,n,r+1,n-1}=S_k\big(w_{n-2}(I_{n-1;r+1}\frac{d}{dt}\ln w_{n-1}-I_{n;r+1})\big).
	\label{prelimexpm}
\end{equation}
Substituting (\ref{prelimexpm}) back in (\ref{smnr}) and down-shifting the obtained expression with respect to $k$, we get
$$
	m_{n-1;n-1,r}=\dot{m}_{n-1;n-1,r+1}-w_{n-2}(I_{n-1;r+1}\frac{d}{dt}\ln w_{n-1}-I_{n;r+1}),
$$
or 
\begin{equation}
	I_{n-1;r}=\frac{\dot{m}_{n-1;n-1,r+1}}{w_{n-2}}-I_{n-1;r+1}\frac{d}{dt}\ln w_{n-1}+I_{n;r+1}.
	\label{prelast1}
\end{equation}
On noting that 
\begin{equation}
	\frac{\dot{m}_{n-1;n-1,r+1}}{w_{n-2}}=\dot{I}_{n-1;r+1}+I_{n-1;r+1}\frac{d}{dt}\ln w_{n-2}
	\label{difint}
\end{equation}
and down-shifting with respect to $r$, we obtain (\ref{recursd1}).\vspace{2mm}

{\it Formula for $t$-integrals.} Our starting point is the relation 
$$
S_k(m_{n-1;p,n-1})=m_{1,p+1,n-1,n}.
$$
Sylvester's identity corresponding to indices $(1,p+1,n-1,n)$ is given by
$$
w_n m_{1,p+1,n-1,n}=m_{1,n-1}m_{p+1,n}-m_{1,n}m_{p+1,n-1}
$$
or equivalently by
\begin{equation}
	w_n m_{1,p+1,n-1,n}=m_{p+1,n}S_k(\dot w_{n-1})-S_k(w_{n-1})\dot m_{p+1,n}.
	\label{fla0}
\end{equation}
The mixed shift-derivative of $w_{n-1}$ present in (\ref{fla0}) can be eliminated by using
(\ref{2dtodacd}) which, for convenience, we write as
\begin{equation}
	S_k(\dot w_n)=\frac{1}{w_n}(S_k(w_n)\dot w_n+w_{n+1}S_k(w_{n-1})).
	\label{fla1}
\end{equation}
Further, the derivative $\dot{m}_{p+1,n}$ can be eliminated by using $(\ref{IJcontd})_2$ written as
\begin{equation}
	\dot m_{pn}=\frac{1}{w_{n-1}}(m_{pn}\dot w_{n-1}+m_{n-1;p,n-1}w_n).
	\label{fla2}
\end{equation}
Substituting 	(\ref{fla1}) and (\ref{fla2}) in (\ref{fla0}) we obtain
\begin{equation}
	m_{1,p+1,n-1,n}=\frac{1}{w_{n-1}}\big(m_{p+1,n}S_k(w_{n-2})-m_{n-1;p+1,n-1}S_k(w_{n-1})\big).
	\label{prelast11}
\end{equation}
On down-shifting with respect to $p$, dividing by $S_k(w_{n-2})$ and re-arranging terms, relation (\ref{prelast11}) 
becomes (\ref{recusdt}).

\subsubsection*{Lattice case.}

Derivation of (\ref{laticerec}) is similar to the one for (\ref{recusdt}) in the semi-discrete case. 
We start with the obvious relation
\begin{equation}
	S^2_{kl}(m_{n-1;p,n-1})=m_{1,p+1,1,n}.
	\label{inrel2}
\end{equation}
The right hand side is involved in the decomposition of $w_{n}$ by means of Sylvester's identity:
$$
\begin{array}{l}
w_{n} m_{1,p+1,1,n}	=m_{11}m_{p+1,n}-m_{1n}m_{p+1,1}\\
\,\quad\qquad\qquad=S^2_{kl}(w_{n-1})m_{p+1,n}-S_k(w_{n-1})S_l(m_{p+1,n})\\
\,\quad\qquad\qquad=w_{n-1} S^2_{kl}(w_{n-1})J_{n;p+1}-S_k(w_{n-1})S_l(w_{n-1} J_{n;p+1}).
\,\end{array}
$$
Next, we can eliminate $J_{n;p+1}$ from the first term on the right using formula $(\ref{ddints})_2$.
This yields the formula
\begin{multline*}
w_{n} m_{1,p+1,1,n}=S_l(J_{n;p+1})\big(w_{n-1} S^2_{kl}(w_{n-1})-S_k(w_{n-1})S_l(w_{n-1})\big)\\
-w_{n}S^2_{kl}(w_{n-1})S_l\Big(\frac{w_{n-2}}{w_{n-1}} J_{n-1;p+1}\Big)
\end{multline*}
which, in turn, can be simplified by using (\ref{lastlattice}). The result is then given by
$$
m_{1,p+1,1,n}=S_l\Big(J_{n;p+1}S_k(w_{n-2})-S_k(w_{n-1})\frac{w_{n-2}}{w_{n-1}}J_{n-1;p+1}\Big).
$$
Making the substitution given by the latter formula into (\ref{inrel2}) and down-shifting the resulting expression with respect to $l$, we obtain
the formula
$$
S_k(m_{n-1;p,n-1})=J_{n;p+1}S_k(w_{n-2})-S_k(w_{n-1})\frac{w_{n-2}}{w_{n-1}}J_{n-1;p+1}
$$
whcih can be easily recognised as equivalent to (\ref{laticerec}).

{\bf Remarks.} Obviously the following pairs of recurrent formulas are identical:  (\ref{recurrent}) and (\ref{recursd1}), (\ref{recusdt}) and (\ref{laticerec}),
where in the former one has to interchange $x$ and $t$. This implies that the corresponding integrals are also identical.

There is an additional interpretation of the recurrent formulas (\ref{recurrent})-(\ref{laticerec}): when they are combined
with the corresponding relations from (\ref{jcont})-(\ref{ddints}), the compatibility condition yields
one of equations (\ref{2dtodac})-(\ref{2dtodacdd}). Hence the combination gives a linear representation (Lax pair) for the 
corresponding equation. 
\subsection{Closed-form expressions}
Recurrent formulas (\ref{recurrent})-(\ref{laticerec}), due to conditions (\ref{iboundary}), can be summed to give closed-form expressions of functions $I$ and $J$ 
and hence the integrals of (\ref{2dtodac})-(\ref{2dtodacdd}). 
It is sufficient to present them only for the continuous and lattice cases since the set of recurrent formulas
in the semi-discrete case is the combination of the former two.

{\bf Continuous case.} It is convenient to re-write formula (\ref{recurrent}) in the form
$$
J_{p+i;p}=J_{p+i-1;p-1}-\frac{w_{p+i-1}}{w_{p+i-2}}\p_x\Big(\frac{w_{p+i-2}}{w_{p+i-1}} J_{p+i-1;p}\Big)
$$
for $i\ge 1$. For $p=1$ this formula yields
$$
J_{i+1;1}=-\frac{w_{i}}{w_{i-1}}\p_x\Big(\frac{w_{i-1}}{w_{i}} J_{i;1} \Big).
$$
Further, seting $p=2,3,\dots$ we obtain
\begin{equation}
	J_{p+i;p}=\sum_{j=1}^p D_{i+j}(J_{j+i-1;j}),
	\label{jrecurs}
\end{equation}
where differential operator $D_{i+j}$ has the form
$$
D_{i+j}=-\frac{w_{i+j-1}}{w_{i+j-2}}\p_x \frac{w_{i+j-2}}{w_{i+j-1}}.
$$
Noting that (\ref{jrecurs}) is a recurrence with respect to index $i$  we make the self-substitution 
to express $J_{p+i;p}$ in terms of $J_{j+1;j}=\p_x \ln w_j$ for some $j$. This yields the closed-form expression
$$
	J_{p+i;p}=\sum_{j_1=1}^p D_{i+j_1}\sum_{j_2=1}^{j_1} D_{i-1+j_2} \dots \sum_{j_{i-1}=1}^{j_{i-2}} D_{2+j_{i-1}} J_{j_{i-1}+1, j_{i-1}}\\[2mm]
$$

{\bf Discrete case.} As in the continuous case we start off by substituting $n=i+1, p=1$ in (\ref{laticerec}) to get
$$
J_{i+1;1}=\frac{w_{i-1}}{w_i}S_k\Big(\frac{w_{i}}{w_{i-1}}\Big) J_{i;1}.
$$
Continuing  with $n=p+i$ where $p=2,3,\dots$ we observe that
$$
J_{p+i;p}=\sum_{j=1}^p S_k^{p-j}\Big(\frac{w_{i+j-2}}{w_{i+j-1}}S_k\big(\frac{w_{i+j-1}}{w_{i+j-2}}\big) J_{j+i-1;j}\Big).
$$
Again, this is a recurrence with respect to index $i$. The difference between continuous and discrete case is that now the expression for $J_{j+1;j}$ is still quite complicated.
This forces us to run the recurrence one step further down where we have $J_{j;j}=1$. Thus the closed-form expression for $J$ is
$$
J_{p+i;p}=\sum_{j_1=1}^p S_k^{p-j_1} D_{i+j_1} \sum_{j_2=1}^{j_1} S_k^{j_1-j_2}D_{i-1+j_2}\dots \sum_{j_{i}=1}^{j_{i-1}} S_k^{j_{i-1}-j_{i}} D_{1+j_{i}}
$$
where
$$
D_{i+j}=\frac{w_{i+j-2}}{w_{i+j-1}}S_k\Big(\frac{w_{i+j-1}}{w_{i+j-2}}\Big).
$$

\section{Non-vanishing case ($\beta\ne 0$)} \label{nonvan}
Here we show that the integrals can be expressed
in terms of functions $I$ and $J$ introduced before.
We employ the approach used to derive integrals of the Liouville equation in Section \ref{Liouv}.
Namely, instead of the original equation we consider its differentiated (differenced) versions 
which have the form of a vanishing determinant. 

Consider the new quantities $\varphi_{n;r}$, $\vartheta_{n;p}$ given by the following expressions.
\begin{enumerate}[]
\item Continuous case:
\begin{equation}
	\varphi_{n;r}=I_{n;r}\p_t \ln w_{n}-I_{n+1;r},\ \ \vartheta_{n;p}=J_{n;p}\p_x \ln w_{n}-J_{n+1;p}.
		\label{intctnv}
\end{equation}
\item Semi-discrete case:
\begin{equation}
		\varphi_{n;r}=I_{n;r}\frac{d}{dt}\ln w_n-I_{n+1;r},\ \ \vartheta_{n;p}=S_k(J_{n;p})+\frac{w_{n-1}}{S_k(w_{n-1})}J_{n;p+1}.
		\label{intsdnv}
	\end{equation}
\item Lattice case:
\begin{equation}
	\varphi_{n;r}=S_l(I_{n;r})+\frac{w_{n-1}}{S_l(w_{n-1})}I_{n;r+1},\ \ \vartheta_{n;p}=S_k(J_{n;p})+\frac{w_{n-1}}{S_k(w_{n-1})}J_{n;p+1}.
	\label{intslatnv}
\end{equation}
\end{enumerate}

\lemma{Quantities $\varphi_{n;r}$, $\vartheta_{n;p}$ satisfy the following relations.
\begin{enumerate}[]
\item Continuous case:	
\begin{equation}
\p_x \varphi_{n;r}=I_{n-1;r}\frac{w_{n-2}\p_t w_n}{w_{n-1}^2},\ \ 	\p_t \vartheta_{n;p}=J_{n-1;p}\frac{w_{n-2}\p_x w_n}{w_{n-1}^2}.
	\label{dfcontnd}
\end{equation}
\item Semi-discrete case:	
\begin{equation}
\ds \Delta_k\varphi_{n;r}=S_k\Big(\frac{w_{n-2}}{w_{n-1}}I_{n-1;r}\Big)\frac{\dot{w}_n}{w_{n-1}},\ \ \dot{\vartheta}_{n;r}=S_k\Big(\frac{w_{n-2}}{w_{n-1}^2}J_{n-1;p}\Big)\Delta_k w_n.
\label{dfsdnd}
\end{equation}
\item Lattice case:	
\begin{equation}
\Delta_k \varphi_{n;r}=S_{kl}^2\Big(\frac{w_{n-2}}{w_{n-1}}I_{n-1;r}\Big)\frac{\Delta_l w_n}{S_l(w_{n-1})},\ \ \Delta_l \vartheta_{n;p}=S_{kl}^2\Big(\frac{w_{n-2}}{w_{n-1}}J_{n-1;p}\Big)\frac{\Delta_k w_n}{S_k(w_{n-1})}.
\label{dfltnd}
\end{equation}

\end{enumerate}
}
The proof of these relations is quite straightforward. In order to prove, say  $(\ref{dfcontnd})_2$, we differentiate $(\ref{intctnv})_2$ with respect to $t$ and use formulas (\ref{jcont}) and (\ref{2dtodac}) to eliminate the $t$-derivatives of $J$ and the mixed derivative of $w_n$.

Proving (\ref{dfsdnd}) is slightly more involved: differentiating the expression for $\vartheta_{n;r}$ and utilising (\ref{IJcontd}) and (\ref{2dtodacd}), we get
$$
\dot{\vartheta}_{n;r}=S_k\Big(\frac{w_{n-2}w_n}{w_{n-1}^2}J_{n-1;p}\Big)+\frac{w_{n-2}w_n}{w_{n-1}S_k(w_{n-1})}J_{n-1;p+1}-\frac{w_n S_k(w_{n-2})}{S_k(w_{n-1}^2)}J_{n;p+1}.
$$
On using formula (\ref{recusdt}) up-shifted with respect to $p$ to eliminate $J_{n;p+1}$, the previous formula simplifies to $(\ref{dfsdnd})_2$.
Further, differencing the expression for $\varphi_{n;r}$ and making use of formulas (\ref{IJcontd}) and (\ref{2dtodacd}) we obtain $(\ref{dfsdnd})_1$.

Following the same strategy in the lattice case, we difference $(\ref{intslatnv})_2$ and use (\ref{ddints}), (\ref{2dtodacdd}) to obtain
\begin{multline*}
\Delta_l \vartheta_{n;p}=S_k\Big(\frac{w_n}{w_{n-1}}\Big)S_{kl}^2\Big(\frac{w_{n-2}}{w_{n-1}}J_{n-1;p}\Big)+\frac{w_n}{S_k(w_{n-1})}S_l\Big(\frac{w_{n-2}}{w_{n-1}}J_{n-1;p+1}\Big)\\
-\frac{w_n}{S_k(w_{n-1})}S^2_{kl}\Big(\frac{w_{n-2}}{w_{n-1}}\Big)S_l(J_{n;p+1}).
\end{multline*}
On eliminating $J_{n;p+1}$ by means of (\ref{laticerec}) up-shifted with respect to $p$, we get $(\ref{dfltnd})_2$. This concludes the proof.

\stepcounter{theorem}
\theorem{The quantities $\varphi_{N;r},\, \vartheta_{N;p}$ 
are the integrals of equation (\ref{det1}) or equivalently of (\ref{2dtodac})-(\ref{2dtodacdd}) with boundary conditions (\ref{ws0}),(\ref{ws}), assuming $\beta\ne 0$.
}
{\bf Proof.} Substituting $n=N$ to (\ref{dfcontnd})-(\ref{dfltnd}) and noting that (\ref{ws}) implies that the derivatives (differences) of $w_N$ vanish hence
the right hand sides of (\ref{dfcontnd})-(\ref{dfltnd}) vanish as well. This concludes the proof.

\section{General solutions}
In this section we carry out calculations similar to the ones done for the continuous case in \cite{leznsolv} 
which is also included here for the completeness of exposition.

The simplest equation (\ref{ws}) corresponds to $N=2$ and $\beta=0$:
\begin{equation}
	\begin{vmatrix}
	 u & u_t \\
	 u_x & u_{tx}
	\end{vmatrix}=0
	\label{w2}
\end{equation}
and is related to the D'Alembert equation $v_{tx}=0$  via the substitution $u=\exp(v)$.
The general solution of (\ref{w2}) is therefore $u=X_1(x)T_1(t)$, where $X_1$ and $T_1$
are arbitrary functions. The form of equations and the solution in the simplest case suggest
trying the ansatz 
\begin{equation}
	u(t,x)=\sum_{i=1}^{N-1}X_i(x) T_i(t)
	\label{gsol0}
\end{equation}
for arbitrary $N$. Note that this ansatz contains the same number of arbitrary functions as the order
of the equation hence it is a good candidate for the general solution.

After substituting (\ref{gsol0}) into matrix $\A$ we can represent it as the product of
two rectangular matrices:
\begin{multline*}
	\begin{bmatrix}
	u & u_t & \dots & u_{t^{N-1}} \\
	u_x & u_{tx} &\dots & u_{t^{N-1}x} \\
	\vdots & \vdots & \ddots & \vdots \\
	u_{x^{N-1}} & u_{tx^{N-1}} &\dots & u_{t^{N-1}x^{N-1}}
	\end{bmatrix}=\begin{bmatrix}
	X_1 & X_2 & \dots & X_{N-1} \\
	X'_1 & X'_2 & \dots & X'_{N-1} \\
	\vdots & \vdots & \ddots & \vdots \\
	X^{(N-1)}_1 & X^{(N-1)}_2 & \dots & X^{(N-1)}_{N-1}
	\end{bmatrix}\\
	\times\begin{bmatrix}
	T_1 & T'_1 & \dots & T^{(N-1)}_1 \\
	T_2 & T'_2 &\dots & T^{(N-1)}_2 \\
	\vdots & \vdots & \ddots & \vdots \\
	T_{N-1} & T'_{N-1} &\dots & T^{(N-1)}_{N-1}
	\end{bmatrix}.
\end{multline*}
According to the Cauchy-Binet formula the determinant of $\A$ is zero and hence (\ref{gsol0}) is the general solution 
of (\ref{det1}) in the vanishing case.

The general solution for a slightly more involved case of $\beta\ne 0$ can 
be derived from the solution of $w_N=0$. Indeed, 
the latter equation is equivalent to $\p^2_{tx} \ln w_{N-1}=0$ (see formula (\ref{2dtodac}) ) hence
\begin{equation}
	w_{N-1}=X(x)T(t).
	\label{wn-1}
\end{equation}
On the other hand, the result of substitution of (\ref{gsol0}) to $w_{N-1}$ can be written
as the product of Wronskians of functions $X_i$ and $T_i$:
$$
w_{n-1}\Big(\sum_{i=1}^{N-1}X_i T_i\Big)=W(X_1,\dots,X_{N-1})(x)\,W(T_1,\dots,T_{N-1})(t).
$$
Taking into account the remarks above and using the easily verifiable fact that $w_{N-1}(XT u)=X^{N-1}T^{N-1} w_{N-1}(u)$, we can write the general solution of the equation
$w_N=\beta$ in the form
\begin{equation}
	u(t,x)=\frac{\beta^\frac{1}{N}\sum_{i=1}^N X_i T_i}{\sqrt[N]{W(X_1,\dots,X_{N})(x)\,W(T_1,\dots,T_{N})(t)}}.
	\label{gsolne0}
\end{equation}
{\bf Remark.} Note that this solution contains two ``extra'' arbitrary functions since the equation is of the
order $2N-2$ while the solution is made of $2 N$ arbitrary functions. The number of functions 
can be reduced by re-parametrisation
$$
\frac{T_i}{T_1}\to  T_{i-1},\ \ \frac{X_i}{X_1}\to  X_{i-1},\ \  i=2,\dots,N.
$$

Following exactly the same reasoning we find general solutions for $\beta=0$ in the semi-discrete and lattice cases. 
The general solutions are correspondingly
\begin{equation}
	u(t,k)=\sum_{i=1}^{N-1} T_i(t) K_i(k),\ \ u(k,l)=\sum_{i=1}^{N-1} K_i(k) L_i(l).
	\label{gensolsddd}
\end{equation}
In order to get solutions for $\beta\ne 0$ in the semi-discrete case we note that 
$$
w_{N-1}\big( T u \big)=w_{N-1}(u) T^{N-1}
$$
and
$$
w_{N-1}\Big(\sum_{i=1}^{N-1} T_i(t) K_i(k)\Big)=W(T_1,\dots,T_{N-1})(t)C(K_1,\dots,K_{N-1})(k),
$$
where $C$ stands for the Casoratian of sequences $K_1,\dots,K_{N-1}$.
Hence the general solution can be written in the form
$$
u(t,k)=\frac{\sum_{i=1}^{N} T_i(t) K_i(k)}{\sqrt[N]{W(T_1,\dots,T_N)(t)}},
$$
where sequences $K_i$ satisfy the condition
\begin{equation}
	C(K_1,\dots, K_N)(k)=\beta.
	\label{casorcond}
\end{equation}
For example, we can choose ${K_1,\dots,K_N}$ to be a linearly independent set of solutions of the linear difference equation
$$
K(k+N)+ c_1(k) K(k+N-1)+\dots c_{N-1}(k) K(k+1)+(-1)^N K(k)=0,
$$
where $c_1(k), \dots, c_{N-1}(k)$ are arbitrary sequences.
It follows from Abel's lemma that the Casoratian of such a set is constant. We only need to ensure that sequences $K_i$
are scaled such that condition (\ref{casorcond}) is satisfied.

Similarly in the lattice case we have
$$
w_{N-1}\Big(\sum_{i=1}^{N-1} K_i(k) L_i(l)\Big)=C(L_1,\dots,L_{N-1})(l)C(K_1,\dots,K_{N-1})(k).
$$
Hence the general solution can be represented in the form
$$
u(t,k)=\sum_{i=1}^{N} K_i(k) L_i(l) ,
$$
where sequences $K_i$ satisfy (\ref{casorcond}) and $L_i$ satisfy a similar condition
$$
C(L_1,\dots,L_N)(l)=1.
$$
Note that general solutions of (\ref{2dtodac})-(\ref{2dtodacdd}) are obtained by simply substitution of the solutions obtained above in minors $w_n$.

\section*{Concluding remarks} 

In this paper we have studied the integrability of certain determinantal equations. 
We have proved their Darboux integrability by constructing recurrent and closed-form formulas for integrals 
and corresponding integrating factors. 
The general solutions of equations associated with vanishing determinant are given explicitly while
in the non-vanishing case they are expressed in terms of solutions of linear ordinary equations.
A connection of the determinantal equations with analogues of the 2D Toda equation allows one to construct 
integrals and solutions for the latter as well. 

Of course there are other systems that can be derived from the determinant.  A few different ways to introduce them 
were already indicated in \cite{demint}. This line of research 
will be continued in a separate publication. Another problem yet to be considered
is enumerating various reductions of the discrete analogues of the 2D Toda system. Since the underlying determinantal
structure for both continuous and discrete equations is the same, then there should be no formal obstacles
for carrying  across the reductions from continuous case.

An important attribute of integrability which is not considered in this paper, is the structure of higher symmetries.
For Darboux integrable equations the higher symmetries can be constructed from integrals. The only currently missing 
ingredient in this construction is the operators that map integrals to symmetries. These operators can be shown closely 
related (see e.g. Section \ref{Liouv}) to the integrating factors constructed in this paper.
\vspace{2mm}

\subsubsection*{Acknowledgements.} Authors are grateful to Ken Russell for attention to this work
and constructive comments on the presentation.


\vspace{5mm}


\begin{thebibliography}{10}

\bibitem{hir0} Hirota R, Ohta, Y and  Satsuma J 1988  Wronskian structures of solutions for soliton equations. Recent developments in soliton theory {\it  Progr. Theoret. Phys. Suppl.}  {\bf  94} 59--72

\bibitem{hir1} Hirota R, Ito  M and Kako F 1988 Two-dimensional Toda lattice equations. Recent developments in soliton theory. {\it Progr. Theoret. Phys. Suppl.}  {\bf  94}  42--58

\bibitem{hir2} Ohta Y, Hirota R, Tsujimoto S and  Imai T 1993 Casorati and discrete Gram type determinant representations of solutions to the discrete KP hierarchy {\it J. Phys. Soc. Japan}  {\bf 62}  1872--1886

\bibitem{darb} G. Darboux 1915 {\it  Le\c{c}ons sur la th\'eorie g\'en\'erale des surfaces et les applications g\'eom\'etrques du calcul infinit\'esimal} (Paris: Hermann)  V.2

\bibitem{demint} Demskoi D K  2010 Integrals of Open Two-dimensional Lattices  {Theor. Math.  Phys.} {\bf  163} 466--471

\bibitem{fordy} Fordy A P and Gibbons J 1980 Integrable Nonlinear Klein-Gordon Equations and Toda Lattices {\it Commun. Math. Phys} {\bf 77}  21--30

\bibitem{gours} Goursat, E  1895 Sur une classe d'\'equations aux d\'eriv\'ees partielles du second ordre, et sur la th\'eorie des int\'egrales interm\'ediaires {\it Acta Math.}  {\bf 19}  285--340

\bibitem{vessiot} Vessiot, E 1942 Sur les \'equations aux d\'eriv\'ees partielles du second ordre, $F(x,y,z,p,q,r,s,t)=0$, int\'egrables par la m\'ethode de Darboux {\it J. Math. Pures  Appl. (9)}  {\bf 21} 1--66

\bibitem{zhsok} Zhiber A V and  Sokolov V V 2001 Exactly integrable hyperbolic equations of Liouville type {\it Russian Math. Surveys } {\bf 56}  61--101

\bibitem{anders} Anderson I M and  Kamran N 1997 The variational bicomplex for hyperbolic second-order scalar partial differential equations in the plane {\it Duke Math. J. } {\bf 87}   265--319

\bibitem{shabyam} Shabat A B and Yamilov R I 1981 Exponential systems of type I and Cartan matrices {\it  preprint}  Bashkirian Branch, USSR Acad. Sci., Ufa

\bibitem{shablez} Leznov A N, Smirnov V G and Shabat A B 1982 Internal symmetry group and integrability conditions for two-dimensional dynamical systems {\it Teoret. Mat. Fiz. }{\bf 51} 10--21 [ in Russian]

\bibitem{leznov} Leznov A N and  Saveliev M V 1985 {\it Group-Theoretical Methods for Integration of Nonlinear Dynamical Systems}
(Nauka, Moscow) [in Russian]; English transl., Birkhauser, Basel (1992).

\bibitem{zhgur} Gur'eva A M and  Zhiber A V 2004 Laplace invariants of two-dimensionalized open Toda chains {\it Teoret. Mat. Fiz.} {\bf 138} 401--421 [ in Russian] ; translation in {\it Theoret. and Math. Phys.} {\bf 138} 338--355

\bibitem{nie} Nie Z 2014 On characteristic integrals of Toda field theories  {\it J. Nonlinear Math. Phys.}  {\bf 21} 120--131

\bibitem{adst}  Adler V \'E and  Startsev S Ya 1999 On discrete analogues of the Liouville equation {\it Theoret. and Math. Phys.} {\bf 121}  1484--1495

\bibitem{habib0} Habibullin I T 2005 Characteristic algebras of fully discrete hyperbolic type equations {\it SIGMA Symmetry Integrability Geom. Methods Appl. 1} {\bf  023} 9 pp
%
\bibitem{kh1} Habibullin I T, Zheltukhin K, Yangubaeva M 2011 Cartan matrices and integrable lattice Toda field equations {\it J. Phys. A: Math. Theor.}  {\bf  44} 465202, 20 pp

\bibitem{habib1} Garifullin R, Habibullin I T and  Yangubaeva M 2012 Affine and finite Lie algebras and integrable Toda field equations on discrete space-time {\it SIGMA Symmetry Integrability Geom. Methods Appl.~8}

\bibitem{smirnov} Smirnov S V  2014 On Darboux integrability of discrete 2D Toda lattices {\it preprint  arXiv:1410.0319}

\bibitem{mokhov} Mokhov O I 2009 Consistency on cubic lattices for determinants of arbitrary orders {\it Tr. Mat. Inst. Steklova 266, Geometriya, Topologiya i Matematicheskaya Fizika. II} 202--217  [in Russian]; translation in 2006 in  {\it Proc. Steklov Inst. Math. } {\bf 266} 195--209

\bibitem{nijh1}  Nijhoff F W and  Walker A J  2001  The discrete and continuous Painlev\'e VI hierarchy and the Garnier systems. Integrable systems: linear and nonlinear dynamics {\it  Glasg. Math. J.} {\bf  43A }109--123.

\bibitem{bob1}  Bobenko A I, Suris Yu B 2002 Integrable systems on quad-graphs {\it Int. Math. Res. Not.}  {\bf  11} 573--611.

\bibitem{bob2}  Adler V E, Bobenko A I and Suris Yu B 2003 Classification of integrable equations on quad-graphs. The consistency approach {\it  Comm. Math. Phys.} {\bf 233} 513--543.

\bibitem{tsar1} Tsarev  S P and  Wolf T 2009 Hyperdeterminants as integrable discrete systems {\it J. Phys. A } {\bf 42}  454023, 9 pp

\bibitem{demstar} Demskoi D K and  Startsev S Ya 2004 On the construction of symmetries from integrals of hyperbolic systems of equations {\it Fundam. Prikl. Mat. } {\bf 10}  29--37 [in Russian] ; translation in {\it J. Math. Sci. (N. Y.) }{\bf 136}  4378--4384

\bibitem{robquisp}  Quispel G R W, Capel H W and  Roberts J A G 2005 Duality for discrete integrable systems {\it J. Phys. A} {\bf  38} 3965--3980.

\bibitem{rebel} Rebelo R and  Valiquette F 2013 Symmetry preserving numerical schemes for partial differential equations and their numerical tests {\it J. Difference Equ. Appl.} {\bf 19}  738--757

\bibitem{winter} Levi D, Martina L  and Winternitz P 2014 Lie-point symmetries of the discrete Liouville equation {\it preprint } http://arXiv/abs/1407.4043

\bibitem{sylvester} Sylvester J J  1862 Sur une classe d'equations differ\'entielles et d'equations aux differences finies d'une forme integrable
{\it Compt. Rend. Acad. Sc. } {\bf 54} 129 

\bibitem{hirot} Hirota R 2003 Determinants and Pfaffians: How to obtain N-soliton solutions from 2-soliton solutions {\it RIMS K\^oky\^uroku } {\bf 1302} 220--242

\bibitem{hirot2} Hirota R 1981 Discrete analogue of a generalized Toda equation {\it J. Phys. Soc. Japan} {\bf  50} 3785--3791.

\bibitem{leznsolv} Leznov A N 1980 On complete integrability of a nonlinear system of partial differential equations in two-dimensional space {\it  Teoret. Mat. Fiz.} {\bf 42}  343--349 [in Russian]

\end{thebibliography}
\end{document}